\documentclass[aps,twocolumn,floatfix]{revtex4} 

\usepackage{amsmath,amsfonts,amssymb,graphicx,color,times,bbm}

\begin{document}

\bibliographystyle{apsrev}
\newtheorem{theorem}{Theorem}
\newtheorem{corollary}{Corollary}
\newtheorem{definition}{Definition}
\newtheorem{proposition}{Proposition}
\newtheorem{lemma}{Lemma}
\newcommand{\proofend}{\hfill\fbox\\\medskip }
\newcommand{\proof}[1]{{\bf Proof.} #1 $\proofend$}
\newcommand{\nn}{{\mathbbm{N}}}
\newcommand{\rr}{{\mathbbm{R}}}
\newcommand{\cc}{{\mathbbm{C}}}
\newcommand{\mbp}{\ensuremath{\spadesuit}}
\newcommand{\je}{\ensuremath{\heartsuit}}
\newcommand{\jd}{\ensuremath{\clubsuit}}
\newcommand{\id}{{\mathbbm{1}}}
\renewcommand{\vec}[1]{\boldsymbol{#1}}
\newcommand{\me}{\mathrm{e}}
\newcommand{\mi}{\mathrm{i}}
\newcommand{\md}{\mathrm{d}}
\newcommand{\sg}{\text{sgn}}

\delimitershortfall=-2pt

\title{Reconstructing quantum states efficiently}

\author{M.~Cramer$^{1,2,}$ and M.~B.~Plenio$^{1,2}$}

\affiliation{${}^1$ Institut f\"ur Theoretische Physik, 
Albert-Einstein Allee 11, Universit\"at Ulm, D-89069 Ulm, Germany}
\affiliation{${}^2$ Institute for Mathematical Sciences, Imperial 
College London, London SW7 2PG, UK}

\begin{abstract}
Quantum state tomography, the ability to deduce 
the density matrix of a quantum system from measured data, is of 
fundamental importance for the verification of present and future 
quantum devices. It has been realized in systems with few components 
but for larger systems it becomes rapidly infeasible because the 
number of quantum measurements and computational resources required 
to process them grow exponentially in the system size. Here we show 
that we can gain an exponential advantage over direct 
state tomography for quantum states typically realized in nature. Based on singular value 
thresholding and matrix product 
state methods we introduce a state reconstruction scheme 
that relies only on a linear number of measurements. The 
computational resources for the postprocessing required to reconstruct 
the state with high fidelity from these measurements is polynomial in 
the system size. 
\end{abstract}

\maketitle

\date{\today}

It is one of the principal features distinguishing classical from 
quantum many-body systems, that for the former the specification of 
a state requires a parameter set whose size scales linearly in the 
number of subsystems, while in the latter this set scales exponentially. 
It is this difference that supports the observation that quantum 
devices appear to be fundamentally hard (exponential in the number
of subsystems) to simulate on a classical computer and may in turn 
possess a computational power exceeding that of classical devices
\cite{Feynman}. 

This presents us with the blessing of being able to construct 
information processing devices fundamentally superior to any 
classical device and the curse of their complexity, challenging our 
ability to verify efficiently that such a quantum information 
processing device or quantum simulator is actually functioning as 
intended. Such devices and, more generally, quantum simulators
 may be viewed as the preparation of elaborate 
quantum states on which we then carry out measurements. Verifying
efficiently that an intended state---the ground or thermal state 
of a physical Hamiltonian for example---has indeed been prepared 
by a quantum information processor or a quantum simulator is 
essential. 

The full determination of the quantum state of a system, that is 
quantum state tomography \cite{Vogel R 89}, can of course be 
achieved -- one simply measures a complete set of observables whose 
expectation values fully determine the quantum state 
\cite{Smithey 93, Blatt, Leibfried 05, James KMW 01, Lvovsky 09}. 
In practice however, this approach is beset with several problems 
when applied to quantum-many party systems.
{\em Firstly}, in quantum state tomography the size of the set of 
measurements scales exponentially with the number of subsystems. 
For moderately sized systems, such as the electronic state of $8$ 
ions \cite{Blatt}, tomography has been 
demonstrated but it rapidly becomes infeasible for larger systems 
thanks both to excessive time required to carry out the measurements 
and because the precision of those measurements has to increase 
exponentially to ensure that a function of the probability amplitudes 
of the state will not return an essentially random result. 
{\em Secondly}, making the connection
between the measurement data on the one hand and the density matrix
of a state best approximating these data on the other will usually 
require classical postprocessing that cannot be executed efficiently 
on a classical computer (see \cite{Blatt}). 
{\em Thirdly}, writing out the full state of a physical system will 
be impossible for more than approximately $40$ spin-1/2 particles 
and approximate representations from which one can extract
expectation values efficiently to high precision need to be 
used.

Here, we address all of the above challenges at the 
same time and demonstrate the efficiency of the proposed approach 
with numerical examples. We present the case of general 
pure quantum states in some detail and outline generalizations to 
mixed states. To pave the way towards the general argument, we 
start the exposition with a discussion of unique ground 
states of local Hamiltonians.

Consider a $k$-local Hamiltonian acting on a set 
$L$ of $N$ spins arranged on some lattice
equipped with the notion of a distance $dist(i,j)$ between sites $i$ and $j$,
\begin{equation}
        \hat{H}=\sum_{i\in L}\hat{h}_i,
\end{equation}
where each Hamiltonian $\hat{h}_i$ acts on spins that are at most 
at a distance $k$ from spin $i$. Let us collect these in the set 
$I_i=\{j\in L\,:\,\text{dist}(i,j)\le k\}$. The non-degenerate 
ground state of a $k$-local Hamiltonian is the state of lowest 
energy and therefore uniquely determined by the expectation values 
of the $\hat{h}_{i}$. Indeed, if there was another state with the 
same expectation values, its energy would be the same, violating 
the uniqueness assumption. For an unknown $k$-local Hamiltonian 
we do not know the $\hat{h}_{i}$ and cannot restrict measurements to 
these observables only. The ground state $|gs\rangle$ is nevertheless 
uniquely determined by all its reductions to the sites 
$I_i$, 
$\hat{\varrho}_i=\text{tr}_{L\backslash I_i}[|gs\rangle\langle gs|]$, 
as these determine {\it all} possible 
expectation values of operators acting on $I_i$, in particular those of
the unknown $\hat{h}_{i}$. Hence, if
an experiment prepares the unique ground state of some unknown $k$-local 
Hamiltonian, we can determine that state fully by measuring the 
$N$ reduced density matrices $\hat{\varrho}_i$. In this setting 
we have hence overcome the first problem mentioned above as the 
state is fully determined by the $\hat{\varrho}_i$ -- so $N$ 
density matrices of size $2^{|I_i|}\times 2^{|I_i|}$, where 
$|I_i|$ is the cardinality of $I_i$. For a nearest neighbour 
Hamiltonian on a $d$-dimensional lattice, e.g., $|I_i|=2d+1$. 
 Using these insights, we will discuss the case 
of general pure states later.

To overcome the second problem, we require an efficient method to
find a pure state $|\psi\rangle$ whose reduced density matrices
$\hat{\sigma}_i=\text{tr}_{L\backslash I_i}[|\psi\rangle\langle\psi|]$ 
coincide with the $\hat{\varrho}_i$. 
The method of choice is  singular value thresholding (SVT) 
\cite{matrix_completion,svthresh} (see the Appendix for technical
details), which has been developed very recently in the context of 
classical {\em compressive sampling} or {\em matrix completion} \cite{Compressed}
and may also be applied to the quantum setting \cite{Kosut, Gross}.  SVT provides a 
recursive algorithm that converges provably towards a low rank 
solution satisfying a set of linear constraints such as the 
requirement to match reduced density matrices.  
SVT converges rapidly to the solution especially so when it 
has low rank. Unfortunately, SVT as originally proposed is not 
scalable as it requires a full representation 
of a matrix of the same size as the density matrix 
describing the system and a singular value decompositions
of this matrix. Hence, both the requirement for
memory and time scale exponentially in the number of sub-systems
and the  straightforward application of SVT is restricted to 
well below $20$
spin-1/2 particles. However, as we will see, a modification of 
the algorithm allows us to overcome this problem.

Denote by $|\phi\rangle$ the unknown target state and by 
$\hat{\varrho}_i$, $i=1,\dots,N$, its reduced density matrices 
as described above. The following modification of the standard 
SVT algorithm (see appendix) will yield a state $|\psi\rangle$ 
whose reduced density matrices $\hat{\sigma}_i$ closely matches 
those of $|\phi\rangle$. Let $\hat{\sigma}_j^\alpha$, $\alpha=x,y,z,0$, 
the Pauli spin matrices acting on site $j$ and by $\hat{P}_k$ 
denote all the possible operators $\prod_{j\in I_i}\sigma_{j}^{\alpha_j}$, 
$i=1,\dots,N$, of which there are $\sum_{i\in L}4^{|I_i|}=:K$. 
From $\hat{\varrho}_i$ we know the expectation values 
$\langle \phi|\prod_{j\in I_i}\sigma_{j}^{\alpha_j} |\phi\rangle
=\text{tr}_{I_i}[\hat{\varrho}_i\prod_{j\in I_i}\sigma_{j}^{\alpha_j}]$
and hence the numbers $p_k=\langle \phi|\hat{P}_k |\phi\rangle$, 
$k=1,\dots,K$. The algorithm may then be described as follows. First 
set up the operator $\hat{R}=\sum_{k=1}^Kp_k\hat{P}_k/2^N$ and 
initialize $\hat{Y}_0$ (e.g., by the zero matrix). Then proceed inductively by finding the 
eigenstate $|y_n\rangle$ with largest eigenvalue, $y_n$, of $\hat{Y}_n$
and set 
\begin{equation}
\hat{X}_n=y_n\sum_{k=1}^K\frac{\langle y_n|\hat{P}_k|y_n\rangle}{2^N}\hat{P}_k,\;
\hat{Y}_{n+1}=\hat{Y}_n+\delta_n(\hat{R}-\hat{X}_n).
\end{equation}
A rigorous proof of convergence of $\hat{\sigma}_i=\text{tr}_{I_i}[|y_n\rangle\langle y_n|]$ to $\hat{\varrho}_i$ (equivalently of $\langle y_n|\hat{P}_k|y_n\rangle$ to $p_k$) will be presented elsewhere. Heuristically, convergence is suggested by the extensive
numerics below and can be 
expected from the fact that SVT possesses a convergence proof for 
small $\delta_n\in\rr$ \cite{svthresh}, see Appendix. 

So far, this algorithm still 
suffers from the fact that in every step the $2^N\times 2^N$ matrix $\hat{Y}_n$ needs to be diagonalized. However, the 
$\hat{Y}_n$ are of the form $\sum_{k=1}^Ka_k\hat{P}_k$, $a_k\in\rr$, i.e., they have the form of a 
local ``Hamiltonian". In one spatial dimension, the ground 
states of local Hamiltonians are well approximated by matrix product 
states (MPS) \cite{FannesNW 92, MPSref}. 
Hence, $|y_n\rangle$ can be determined employing
MPS algorithms \cite{Rommer 95,Schollwoeck}, for which the number of parameters 
scale polynomially in the system size and converge rapidly 
\cite{Schuch,Aharonov}. Hence, for one spatial dimension,
we have overcome the second and third problem mentioned above:
This postprocessing is efficient as MPS algorithms are and a MPS provides
an efficient representation of the state as it depends only on linearly many parameters.
Any general pure state may be represented by a MPS and generic MPS are unique
ground states of local Hamiltonians \cite{MPSref}. Hence, the above algorithm
will produce a state that is close to the target state if the MPS dimension and 
the size of the reduced density matrices is chosen sufficiently large.

For higher spatial dimensions MPS are not 
 efficient representations and for optimal performance 
 they need to be replaced by other variational classes. A variety 
 of MPS generalization have been proposed of which the most 
 promising are perhaps the tensor-tree ansatz \cite{TTN} and 
 MERA-approach \cite{Mera}, PEPS \cite{PEPS} and weighted graph 
 states \cite{Anders06}. For each of these, numerical algorithms
 have been developed that determine the largest eigenvalue of a 
 local Hamiltonian. Being the key ingredient in our modified
 SVT method developed here, these more general variational
 classes may be combined naturally in the way we have described 
 for MPS.

\begin{figure*}
\includegraphics[width=1.9\columnwidth]{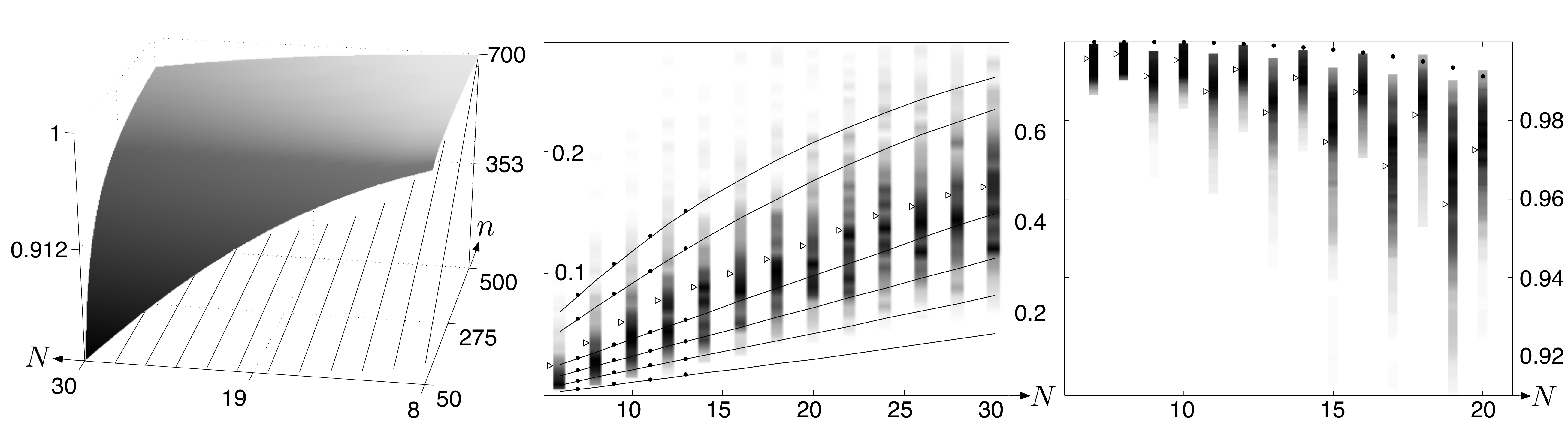}
\caption{\label{fig1} Fidelity $f_{N,n}=|\langle \phi|y_n\rangle|^2$ as a function of the number of spins $N$ and iterations $n$ of the MPS-SVT algorithm for different target states $|\phi\rangle$. Left: Ground state of the locally rotated critical Ising model, $f_{N,n}$ (surface, left axis) and $1/(1-f_{N,n})$ (lines, right axis), showing that, for fixed system size, $1-f_{N,n}$ decreases as $\sim \!1/n$. Middle: $\sqrt{1-f_{N,n}}$ as a function of $N$
for the ground state of the locally rotated critical Ising model (lines, right axis, from top to bottom $n=5, 10, 50, 100, 200, 500$, dots are obtained by exact numerical diagonalization of the Hamiltonian and the $\hat{Y}_n$) and random Hamiltonians (1000 for each $N$) as described in the main text after $n=5$ iterations (densities, left axis, arrows indicate the mean). In both cases the scaling for fixed $n$ of $1-f_{N,n}$ is better than $\sim \!N^2$. Right: $W$ state 
 as described in the text for 4000 MPS-SVT iterations. Plot shows $|\langle \phi|y_n\rangle|^2$ as a function of the number of ions, $N$, for no noise (dots) and Gaussian noise (densities obtained from 100 realizations for each $N$, arrows indicate mean) with a standard deviation of $0.005$ (even $N$) and $0.01$ (odd $N$).
}
\end{figure*}

Let us now consider numerical examples for different target states 
$|\phi\rangle$  to demonstrate the feasibility and 
efficiency of the proposed algorithm. We start with ground
states of nearest-neighbor Hamiltonians on a chain, i.e., the 
$|\phi\rangle=|gs\rangle$ are completely determined by all the reductions to two adjacent 
spins as outlined above and the above algorithm not only produces states that match 
the reduced density matrices of the ground states but, in fact, 
states that are themselves close to the ground states.
 Among ground states of one-dimensional nearest-neighbor Hamiltonians the critical 
ones are the most challenging to approximate by MPS as 
they violate the entanglement-area law \cite{EisertRMP} and we 
test our algorithm for such an example: the critical Ising model. In order for the local operators 
in the Hamiltonian not to be 
exactly the ones that are measured, we also rotate the Ising model 
locally by $\pi/4$ around each spin axis. 
This model is solvable and in order to show 
that we do not consider a pathological case, we also consider 
one-dimensional random Hamiltonians of the form
\begin{equation}
 \hat{H}=\sum_{i=1}^{N-1}\hat{r}^{(i)}_i\hat{r}^{(i)}_{i+1},
\end{equation}
where the $\hat{r}^{(i)}_i$, $\hat{r}^{(i)}_{i+1}$ act on spin 
$i$ and $i+1$, respectively, and are hermitian matrices with 
entries that have real and imaginary part picked from a uniform 
distribution over $[-1,1]$. For each Hamiltonian, we first 
determine the ground state $|gs\rangle$ exactly (i.e., the target state
$|\phi\rangle$) and its reductions and then computed
the fidelity $|\langle gs|y_n\rangle|^2$ after $n$ iterations 
of the MPS-SVT algorithm, see Fig.\ \ref{fig1}.

Our method is of interest for all situations in which standard tomography will not be feasible. This
is the case for the verification of state preparation in 
experiments with too many particles. An example is the recent ion trap experiment
\cite{Blatt} for the preparation of W-states,
$|\phi\rangle=(|10\cdots 0\rangle+|010\cdots 0\rangle+\cdots+|0\cdots 01\rangle)/\sqrt{N}$,
that were limited 
to 8 qubits principally because the classical postprocessing of 
data became prohibitive for longer chains. Here we demonstrate
the efficiency of our approach (we are not limited to few ions and demonstrate
convergence for up to $20$ ions -- even higher number of ions are easily
accessible due to the MPS alteration of
the SVT method) by illustrating how one would 
postprocess experimentally obtained reduced density matrices 
to guarantee the generation of $|\phi\rangle$ or a state very 
close to it. We mimic experimental noise by adding Gaussian 
distributed random numbers with zero mean to the $p_k$. After 
initializing the MPS algorithm with the MPS representation of 
$|\phi\rangle$ and $\hat{Y}_0=\hat{R}$, we use 
$x_n:=\sum_k|p_k-\langle y_n|\hat{P}_k|y_n\rangle|$ as a figure 
of merit for convergence, i.e., after a given number of iterations, 
we pick the $|y_n\rangle$ with minimum $x_n$. The result of such 
a procedure is shown in Fig.\ \ref{fig1}.

So far we have presented the method for pure states
and one-dimensional systems. The SVT algorithm as described above
works for higher-dimensional systems as well and may be made
efficient by adopting higher dimensional analogues of MPS methods
as outlined above. The extension to mixed states is also straightforward as its 
treatment can be reduced to that of pure states by using the fact 
that every mixed state on $N$ qubits can be purified to a pure 
state on $2N$ qubits. Hence we may ask for a globally pure state
on $2N$ qubits that matches the reduced density on all contiguous 
sites of $k$ qubits on the first $N$ qubits. While the reduced 
density matrices do not uniquely determine the mixed state, 
approximations of better and better quality can be obtained 
by increasing $k$. As an example, suppose the state is the Gibbs 
state corresponding to a $k$-local Hamiltonian $\hat{H}$, i.e., the 
state $\hat{\varrho}$ minimizing the free energy
\begin{equation}
\text{tr}[\hat{\varrho}\hat{H}]-TS(\hat{\varrho}).
\end{equation} 
The first term is, as before, for a $k$-local Hamiltonian,
determined by the reduced density matrices. The entropy of 
the total state however can only be learnt exactly from the 
complete density matrix. However, for essentially
all reasonable physical systems, the entropy density 
$\lim_{k\rightarrow\infty} S(tr_{k+1,...}(\rho))/k$ 
in the thermal state of a Hamiltonian exists \cite{BratelliRobinson} 
and as a consequence the total entropy of the state can be 
estimated efficiently from the knowledge of reduced density matrices.

Our algorithm described above may also be adapted straightforwardly
to determine hypothesis states in recently proposed algorithms for 
quantum learning \cite{Aaronson 06}. Here, a small given set of randomly
chosen observables is measured and on the basis of the
measurement outcomes a quantum state closely approximating the
measured expectation values needs to be found. This state will, with
large probability, predict the expectation values of all observables. 
Present approaches to determine such states are based on semi-definite 
programming and are therefore inherently non-scalable, they are
limited to perhaps $12$ qubits \cite{Aaronson 06}. 

Our algorithm will be essential for efficient tomography and 
verification of medium to large scale quantum information devices. 
Already today they are beginning to reach scales for which standard 
tomography is not feasible anymore. Furthermore, it 
may also be applied to problems in condensed matter physics where 
the system has too many components to achieve tomography by standard 
means. 

Hence our combination of singular value thresholding with the matrix
product state representation is expected to become a
useful tool in a wide variety of physical settings and algorithms. 
 
\acknowledgements
This work has been supported by the EU Integrated Project 
QAP, the STREP HIP and an Alexander von Humboldt 
Professorship. The authors acknowledge discussion with
F.G.S.L. Brand{\~a}o at early stages of this project.

\begin{center}
{\bf Appendix}
\end{center}

In singular
value thresholding \cite{matrix_completion,svthresh}
one seeks a solution for the minimization of the trace 
norm of a matrix $X$ subject to some linear constraints, 
i.e.
\begin{eqnarray*}
        && \mbox{minimize}\; \text{tr} |X| \\
        && \mbox{subject to}\; {\cal P}_{\Omega}(X) = 
        {\cal P}_{\Omega}(M)
\end{eqnarray*}
where ${\cal P}_{\Omega}(M)$ is a matrix whose
elements are non-zero only on the entries belonging 
to the index set $\Omega$. For a value $\tau>0$
and a sequence $\{\delta_k\}_{k \ge 1}$ one inductively
defines
\begin{eqnarray*}
        && X^{(k)} = \mbox{shrink}(Y^{(k-1)},\tau)\\
        && Y^{(k)} = Y^{(k-1)} +\delta_k {\cal P}_{\Omega}(M-X^{(k)}),
\end{eqnarray*}
where, in standard SVT, 
\begin{equation}
        \mbox{shrink}(Y,\tau) = U \text{diag}(\{\max\{0,\sigma_i-\tau\}\})
        V^{\dagger}
\end{equation}
with the singular value decomposition 
$Y= U\text{diag}(\{\sigma_i\})V^{\dagger}$. 

If it is our goal to reconstruct pure states compatible with
given reduced density matrices we may adapt SVT and in the 
process make it suitable for application to matrix product 
states. To this end we introduce a small but crucial variation 
of the shrink operation. Rather than introducing a threshold 
$\tau$ we retain only the largest singular value $\sigma_1$
and the corresponding matrix $U\text{diag}(\max_i\sigma_i\,0\cdots 0) V^{\dagger}$. If the target 
state is pure, i.e., a matrix with rank equal to one, this can be expected to converge rapidly, an 
expectation that is confirmed by extensive numerics.

While ${\cal P}_{\Omega}(M-X^{(k)})$ is not itself positive,
initializing the recursion with a positive operator, e.g. a 
pure state, and choosing sufficiently small $\delta_k$ will 
ensure that in the second step of the recursion a matrix is 
generated whose largest eigenvalue is positive and equal
to the largest singular value. 

Crucially, this largest eigenvalue and corresponding eigenstate can then be computed efficiently
via the maximization of the expectation value of a matrix product state 
 solving the multi-quadratic optimization 
problem by a succession of quadratic optimization problems each 
of which can be solved via a generalized eigenvalue problem 
\cite{Rommer 95}. Furthermore, ${\cal P}_{\Omega}(M-X^{(k)})$ in the recursion is
a sum of a linear number of Paulistrings whose expectation
values in a matrix product state may be obtained efficiently.

Reduced density matrices are obtained by measuring the set of strings of Pauli operators $\hat{P}_k$, $k=1,\dots, K$. Then
${\cal P}_{\Omega}(Y) = \sum_{k=1}^K 
\text{tr}[Y\hat{P}_k]\hat{P}_k/2^N$ 
and is hermitean.


\begin{thebibliography}{99}

\bibitem{Feynman} R.P.\ Feynman, Int.\ J.\ Theo.\ Phys.\ {\bf 21},
467 (1982).

\bibitem{Vogel R 89} K.\ Vogel and H.\ Risken, Phys.\ Rev.\ A {\bf 40},
2847 (1989).

\bibitem{Smithey 93} D.T.\ Smithey, M.\ Beck, M.G.\ Raymer, A.\ Faridani,
Phys.\ Rev.\ Lett.\ 
{\bf 70}, 1244 (1993).

\bibitem{Blatt} H.\ H{\"a}ffner, W.\ H{\"a}nsel, C.F.\ Roos, 
J.\ Benhelm, D.\ Chek-al-kar, M.\ Chwalla, T.\ K{\"o}rber, U.D.\ Rapol, 
M.\ Riebe, P.O.\ Schmidt, C.\ Becher, O.\ G{\"u}hne, W.\ D{\"u}r, and R.\ Blatt,
Nature {\bf 438}, 643 (2005).

\bibitem{Leibfried 05} D.\ Leibfried, E.\ Knill, S.\ Seidelin, J.\ Britton, 
R.B.\ Blakestad, J.\ Chiaverini, D.B.\ Hume, W.M.\ Itano, J.D.\ Jost, 
C.\ Langer, R.\ Ozeri, R.\ Reichle, and D.J.\ Wineland, Nature {\bf 438}, 639 (2005).

\bibitem{James KMW 01} D.F.V.\ James, P.G.\ Kwiat, W.J.\ Munro, 
and A.G.\ White, Phys.\ Rev.\ A {\bf 64}, 052312 (2001).

\bibitem{Lvovsky 09} A.I.\ Lvovsky and M.G.\ Raymer, Rev.\ 
Mod.\ Phys.\ {\bf 81}, 299 (2009).

\bibitem{matrix_completion} E.J.\ Candes and B.\ Recht, arXiv:0805.4471 [cs.IT].

\bibitem{svthresh} J-F.\ Cai, E.J.\ Candes, and Z.\ Shen, arXiv:0810.3286 [math.OC].




\bibitem{Compressed} E.J.\ Candes and M.B.\ Wakin, IEEE Sig.\ Proc.\ Mag.\ {\bf 25}, 21 (2008).

\bibitem{Kosut}         R.L.\ Kosut, arXiv:0812.4323 [quant-ph].

\bibitem{Gross} D.\ Gross, Y-K.\ Liu, S.T.\ Flammia, S.\ Becker, and J.\ Eisert, arXiv:0909.3304 [quant-ph];
D.\ Gross, arXiv:0910.1879 [cs.IT].




\bibitem{FannesNW 92} M.\ Fannes, B.\ Nachtergaele, and R.F.\ Werner,
Comm.\ Math.\ Phys.\ {\bf 144}, 443 (1992).

\bibitem{MPSref} D.\ Perez-Garcia, F.\ Verstraete, M.M.\ Wolf, and
J.I.\ Cirac, Quant.\ Inf.\ Comp.\ {\bf 7}, 401 (2007).

\bibitem{Schollwoeck} U.\ Schollw{\"o}ck, Rev.\ Mod.\ Phys.\ {\bf 77}, 259 (2005).

\bibitem{Rommer 95} S.\ {\"O}stlund and S.\ Rommer, Phys.\ Rev.\ Lett.\ 
{\bf 75}, 3537 (1995).


\bibitem{Schuch} N.\ Schuch and J.I.\ Cirac, arXiv:0910.4264
[quant-ph].

\bibitem{Aharonov} D.\ Aharonov, I.\ Arad, and S.\ Irani, arXiv:0910.5055 [quant-ph].

\bibitem{TTN} Y.\ Shi, L.\ Duan and G.\ Vidal, Phys.\ Rev.\ A {\bf 74}, 022320 (2006).
%
\bibitem{Mera} M.\ Rizzi, S.\ Montangero, and G.\ Vidal, Phys.\ Rev.\ A {\bf 77}, 052328 (2008).
%
\bibitem{PEPS} F.\ Verstraete and J.I.\ Cirac, arXiv:cond-mat/0407066.

\bibitem{Anders06} S.\ Anders, M.B.\ Plenio, W.\ D{\"u}r, F.\ Verstraete,
and H.J.\ Briegel,
Phys.\ Rev.\ Lett.\ {\bf 97}, 107206 (2006).

\bibitem{EisertRMP} J.\ Eisert, M.\ Cramer, and M.B.\ Plenio,
Rev.\ Mod.\ Phys.\ {\bf 82}, 277 (2010). 




\bibitem{BratelliRobinson} O.\ Bratteli and D.W.\ Robinson,
{\em Operator Algebras and Quantum Statistical Mechanics}, 
Texts and Monographs in Physics Vol.\ 2 (Springer, New York)
1979.

\bibitem{Aaronson 06} S.\ Aaronson, arXiv:quant-ph/0608142.

\end{thebibliography}
\end{document}